# Challenges and New Directions in Augmented Reality, Computer Security, and Neuroscience

## Part 1: Risks to Sensation and Perception

June 27, 2018

**Stefano Baldassi** (Analytics and Neuroscience, Meta Company), s.baldassi@metavision.com
**Tadayoshi Kohno** (Paul G. Allen School of Computer Science & Engineering, University of Washington), yoshi@cs.washington.edu
**Franziska Roesner** (Paul G. Allen School of Computer Science & Engineering, University of Washington), franzi@cs.washington.edu
**Moqian Tian** (Analytics and Neuroscience, Meta Company), m.tian@metavision.com

*\*The Authors have contributed equally to this manuscript and are thus listed in alphabetical order.*

# 1. Introduction and Motivation

Emerging Augmented Reality (AR) technologies display three-dimensional content (e.g., images or sound) overlaid on a user's perception of the physical world. This content is laid out around a user in the same spatial coordinates as the physical objects surrounding her or him, and with interaction techniques that approximate interactions with physical-world objects (e.g., grabbing and moving them). AR technologies are becoming increasingly sophisticated, with examples appearing in head-mounted displays (HMDs) and on automotive windshields.

With its augmentation of the physical world, AR offers the possibility to benefit users as they perform real-world tasks. For example, manufacturers can superimpose assembly instructions on the physical artifact they are building, showing what action to perform next; surgeons can have 3D scans of a patient's body overlaid on the actual part of the body they describe; or soldiers in the battlefield can have information about the ongoing mission and markers to help spot key landmarks in the field (as those seen guiding the fictional character Tony Stark in the Iron Man movies). Today, these technologies are no longer science fiction but rapidly becoming commercially available.

While AR technologies can and will provide numerous positive benefits to users, history has taught us that as technologies mature, those technologies can become more attractive to computer security adversaries.[1,2] Adversaries might author programs that appear in an AR technology's App Store, or they might exploit a vulnerability in an AR application and thereby cause it to misbehave. To minimize the potential impact of these adversaries in the future, it is important and valuable to consider what undesirable things *might* happen if adversaries manifest in AR environments, and to do so *before* the adversaries manifest. By considering possible adversaries early, it is possible for the AR and computer security communities to take proactive steps to mitigate against possible future threats.

The computer security community has already started to consider several classes of computer security risks for AR technologies. For example, prior works consider privacy implications with an AR technology's use of continuous sensors, like cameras, as well as mechanisms to minimize undesirable privacy exposures.[3,4,5,6] Other prior works consider methods to regulate an AR technology's output,[7,8] to minimize the likelihood of an AR HMD directly blocking a user's perception of the physical world[9] or overlaying deceptive or distracting content (e.g., obscuring or modifying a real-world road sign). A recent paper surveys the broader space of research in security and privacy for AR.[10]

**Our Focus: Sensory and Perceptual Impact.** The above-cited works consider security risks with AR technologies, but do so from a largely computer science perspective and do not consider the intimate relationship between AR technologies and the user's brain. However, as AR technologies become more sophisticated, they will engage with the brain in a way that is much deeper than traditional computing devices.[11] AR technologies offer a much deeper perceptual representation of space and engagement of sensory-motor mechanisms than traditional computer platforms. In AR, the user interface is deployed in the 3D space around the user and can allow direct manipulation of digital content along with the physical content that shares the user's workspace. With the power and sophistication of emerging and future AR technologies comes the possibility of a compromised or malicious app introducing digital content into the user's rich, multi-sensory environment, and thereby interfering with a user's perceptual and cognitive performance,creating potential short- and long-term effects in the user's perception, cognition and motor responses.

---

[1] Y. Zhou and X. Jiang. "Dissecting Android Malware: Characterization and Evolution." In *Proceedings of the IEEE Symposium on Security and Privacy*, 2012.
[2] M. Antonakakis et al. "Understanding the Mirai Botnet." In *Proceedings of the USENIX Security Symposium*, 2017.
[3] S. Jana, D. Molnar, A. Moshchuk, A. M. Dunn, B. Livshits, H. J. Wang, and E. Ofek. "Enabling fine-grained permissions for augmented reality applications with recognizers." In *Proceedings of the USENIX Security Symposium*, 2013.
[4] F. Roesner, D. Molnar, A. Moshchuk, T. Kohno, and H. J. Wang. "World-driven access control for continuous sensing." In *Proceedings of the ACM Conference on Computer & Communications Security*, 2014
[5] L. S. Figueiredo, B. Livshits, D. Molnar, and M. Veanes. "PrePose: Security and privacy for gesture-based programming." In *Proceedings of the IEEE Symposium on Security and Privacy*, 2016.
[6] R. Templeman, M. Korayem, D. Crandall, and A. Kapadia. "PlaceAvoider: Steering first-person cameras away from sensitive spaces." In *Proceedings of the Network and Distributed Systems Security Symposium*, 2014.
[7] K. Lebeck, T. Kohno, and F. Roesner. "How to Safely Augment Reality: Challenges and Directions." In *Proceedings of the 17th Workshop on Mobile Computing Systems and Applications*, 2016.
[8] K. Lebeck, K. Ruth, T. Kohno, and F. Roesner. "Securing Augmented Reality Output." In *Proceedings of the 38th IEEE Symposium on Security and Privacy*, 2017.
[9] E. E. Sabelman and R. Lam. "The Real-Life Dangers of Augmented Reality." In IEEE Spectrum, 2015.
[10] J. A. De Guzman, K. Thilakarathna, and A. Seneviratne. "Security and Privacy Approaches in Mixed Reality: A Literature Survey." arXiv:1802.05797.
[11] S. Baldassi. "The Renaissance of the Third Dimension." https://www.youtube.com/watch?v=HlnhCdujYco&app=. Talk at TedxLA, December 2016.

This paper makes two key contributions:
1. First, we initiate a line of inquiry around the hitherto unexplored area of *potential threats from augmented reality to a user's perception, cognition, and motor responses.*
2. Second, we provide an *initial framework for evaluating these potential threats* according to a set of criteria we develop (e.g., type, longevity, and selectivity of the sensory and perceptual impact).

We encourage AR and computer security researchers as well as cognitive neuroscientists to critically evaluate the spectrum of threats within this framework, as well as the framework itself. We further encourage proactive consideration of these risks by all relevant stakeholders -- neuroscientists, computer security experts, researchers, and industry -- before any adversaries manifest.

**Context.** Our work is squarely at the intersection between AR, computer security, and neuroscience. To conduct this research, we composed a team of experts from all three communities. Our framework reflects our understanding of neuroscience, perception, and computer security threat modeling. For example, much of our discussions derive from a long line of literature in perceptual psychology and neuroscience, with a key difference being that in our work we consider the adversarial uses and extensions of those prior results. Note the views in this paper reflect those of the authors' academic perspective, and not of the organizations they work for.

## 2. Case Studies: Sensory and Perceptual Risks from AR

We begin by presenting several hypothetical case studies of *potential* perceptual issues that may arise due to bugs, accidents, or explicit malice in AR applications or technologies. In the subsequent section, we then take a computer security perspective and discuss *how* such attacks might manifest in practice. Later, we use these case studies to drive the development of a conceptual framework for perceptual AR attacks.

**Background: On Sensory and Perceptual Phenomena.** Human sensory and perceptual systems are the portals through which we experience the world and build our reality. We define as ***Sensory*** all the processing occurring peripherally, at the site of the sensors (e.g., the retina in the eye, the cochlea in the ear). We define as ***Perception*** the processes that transform sensations into information as the neural signal reaches the brain -- in particular the cerebral cortex, where these complex operations will eventually deliver the final experience and the performance that follows. Most of the human decisions are made in real time based on our sensation and perception. The correct position and functioning of our bodies in space and the efficacy of our actions depend on the input we collect and integrate through our senses and our perceptual systems.

As we explore in this paper, AR systems can potentially deliver systematic distortions to the information we perceive. Because of the immersive nature of these technologies and the fact that they engage with our sensory, motor, and cognitive systems in a much deeper way than traditional interfaces, any perturbation to the way people perceive the world may have consequences on behavior that are deeper, and potentially more dramatic, than other traditional technologies (e.g., smartphones). Moreover, because wearable AR is best applied as an augmentation of our capacity to access relevant information and perform in a variety of demanding real world tasks (e.g., surgery, military operations, assembly, navigation, etc.), it is critical to understand in detail how these tasks can be affected by altering the experience of wearable AR users.

**Classes of Phenomena.** In our example case studies below, we will talk in more detail about three different phenomena, or classes of phenomena: ***Photosensitive Epilepsy***, which is a simple paradigmatic case of how the delivery of sensory stimulations can generate harm to users; the vast class of sensory and perceptual ***Adaptation*** phenomena, in which dosed exposures to some stimuli impact on the perception of other stimuli; and some effects and illusions in the domain of ***Motion*** perception, as AR interfaces and their background are in continuous motion and users need to interpret these motion signals correctly to safely interact with the physical world.

These three case studies are by no means exhaustive of the potential threats. Rather, they serve to give an initial taste of the potential impact of introducing malicious sensory stimuli to users via a wearable (e.g., head-mounted display) or other immersive device (e.g., an automotive AR windshield).

## Case Study 1: Photosensitive Epilepsy (PSE)

On December 16, 1997 the Japanese television station TV Tokyo aired an episode of the cartoon "Pocket Monster".[12] That evening, 685 children in Tokyo were reported to have been seen by physicians and about one quarter of them were admitted to the hospital. The visual content of that specific episode contained a flickering pattern that generated convulsions, headache, nausea, general malaise, eye irritation with blurred vision, and seizures. Most of these children had experienced *photosensitive epilepsy (PSE)*, a form of stimulus-induced epilepsy that has an incidence of up to 10% of all new cases of epilepsy detected in the 7-19 years old range.[13] Several other reports can be found about video games, pinball machines, or television shows as triggers for this condition. Though some such exposures are accidental, adversaries have, in the past, deliberately attempted to cause harm to people with photosensitive epilepsy, e.g., in 2008 adversaries placed flashing images on the Epilepsy Foundation website[14] and in 2016 a journalist was targeted with a tweet that successfully induced an epileptic seizure.[15]

---

[12] http://www.cnn.com/WORLD/9712/17/video.seizures.update/

**Risk in AR.** AR technologies could be leveraged by adversaries to trigger PSE. What is very compelling is the simplicity of this trigger. High contrast visual stimuli flickering at temporal frequencies between 10 and 20 Hz are the most likely patterns to activate PSE. The number of people that could be impacted by this is quite limited, but we believe this is a paradigmatic and clearly documented potential harm that can be introduced in a relatively easy fashion by the images displayed with AR technologies. While it is unknown what the full impact of an immersive and wide field of view AR system might have on the delivery of a PSE trigger, we argue that it is important to consider and address this and similar risks before any threats manifest.

## Case Study 2: Adaptation

Sensation and perception are incredibly complex systems in the human brain. One of the most compelling features comes from the fact that our sensory systems adjust their response ("gain") dynamically to adapt to the context they are operating in. *Adaptation and aftereffects*[16] are changes in the sensitivity of our sensory and perceptual responses induced by some form of sustained exposure to specific visual stimuli. For example, when we spend some time in a place with a low luminance level, say our dimly lit bedroom, we become adapted to its luminance range, and it will take a while and a little visual distress to re-adapt after we switch the light on. This well-known visual phenomenon occurs in a comparable fashion in the auditory domain.[17]

Adaptation phenomena are pervasive in our perceptual and cognitive behavior, and they occur at many different levels of the processing pipeline. The above examples (light and audio adaption) are examples of low-level forms of adaptation. However, adaptation occurs and has surprisingly similar manifestations in higher-level aspects of our perceptual cognition, such as for the perception of faces and facial expressions, biological motion, numerosity (e.g., estimating the size of a crowd), affordance (e.g., matching objects with the action towards it), and navigation (moving around in the environment), and so forth.[18]

---

[13] J. A. Quirk, D. R. Fish, S. J. Smith, J. W. Sander, S. D. Shorvon, P.J. Allen. "Incidence of photosensitive epilepsy: a prospective national study." In *Electroencephalography and Clinical Neurophysiology*, 95(4): 260-7, 1995.
[14] https://www.wired.com/2008/03/hackers-assault-epilepsy-patients-via-computer/
[15] https://www.nytimes.com/2017/03/17/technology/social-media-attack-that-set-off-a-seizure-leads-to-an-arrest.html
[16] D. Burr and P. G. Thompson. "Visual aftereffects." In *Current Biology*, 19: R11-R14, 2009.
[17] I. Dean, N. S. Harper, D. McAlpine, and B. L. Robinson. "Rapid neural adaptation to sound level statistics." In *The Journal of Neuroscience: The Official Journal of the Society for Neuroscience*, 28(25): 6430-8, 2008.

Moreover, adaptation is observable in non-visual sensory modalities and in the motor system. Particularly relevant and striking is the case of prism adaptation, in which people who see the world through a prism for long enough will experience severe perceptual-motor disconnections that are coherent with a sensory-motor remapping to match the optical distortions of the world. For example, a person's brain will, after some time, adapt to an upside-down prism by remapping his or her actions to this flipped world. In an excerpt of the TV show "Brain Games", Prof. Daniel Simons show how basketball players are impacted by this effect.[19]

Of particular interest to our article is not only that adaptation manifests itself across most of the sensory and motor domains, but also that its effect can be observed at many different timescales depending on the adapting stimuli. For example, when the prism distortion described above is removed (e.g., a person's perception of the world is returned to rightside-up), it takes time to re-adapt. As another example, the well-known McCollough effect[20] -- a visual effect in which staring at one image for a period of time causes a person to see colors in an otherwise black-and-white image of a specific type -- is known to impact a person's false perception of colors in that black and white image for hours to months, depending on the adaptation schedule of the experimental protocol.[21]

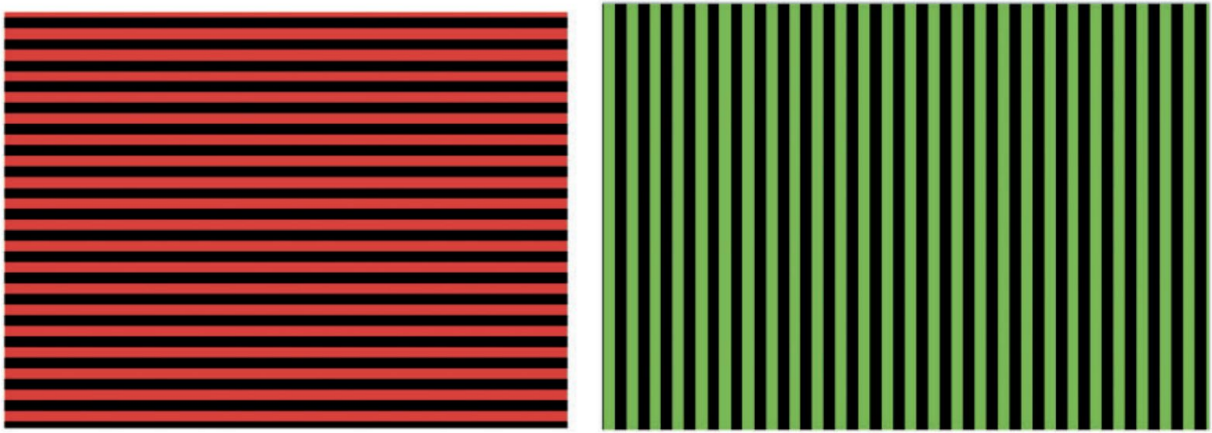

*Figure 1: Images for inducing McCollough Effect (adapted from Ramachandran and Marcus, 2017[22]).* **Warning: this effect can be long-lasting (hours to months).**

---

[18] M.A. Webster. "Visual Adaptation." In *Annual Review of Vision Science*, 2015.
[19] https://www.youtube.com/watch?v=eA2d1tKNFoU
[20] McCollough, Celeste (3 September 1965). "Color Adaptation of Edge-Detectors in the Human Visual System". *Science*. **149** (3688): 1115–1116.
[21] Holding, D.H., & Jones, P.D. "Extremely long-term persistence of the McCollough effect." In *Journal of experimental psychology. Human perception and performance*, 1(4): 323-7, 1975.
[22] Ramachandran, V. S., Marcus, Z. (2017) "Synesthesia and the McCollough effect." *i-Perception* 8: 1–6. doi: 2041669517711718.

**Risk in AR.** Specific "doses" of adapting stimuli, delivered by a malicious or compromised AR application, may similarly affect a user's perception for either a short or long (or possibly permanent) period of time. Such effects could even be done in a way that the user is unaware of them -- like, for example, a subtle shift of the distribution of chromatic profile in the display. Imagine military operations in the field, where small shifts of patterns and color make the difference between your fellow soldiers and your enemies in a generally camouflaged visual scene. "Hacking" the color pattern of the Head-Up Display may, for example, dramatically impact the user's color perception and, consequently, the crucial decision making that relies on color and pattern discrimination.

## Case Study 3: Motion, Sensory Transients, and Motion-Induced Blindness

A flying insect in the visual periphery captures our attention even when we are intensely engaged in another activity. So do transient, sudden stimuli like sharp loud sounds, flashes, or objects that hit us. An incredible computational design in the brain ensures that we process all the motion signals that we sense and use those signals in the proper coordinate system to allow us to navigate environment, avoid obstacles, estimate time to contact, and pay attention to potential dangers.[23]

**Risk in AR.** Moving and temporally modulated stimuli can be used maliciously, to disrupt a user during the execution of delicate tasks. For example, "malicious" motion signals delivered through the visual and the auditory channels could act as stimuli to distract from the task, or to mask and distort perception of task-relevant motion signals from the physical world. Ultimately, such an attack could even perturb postures and locomotion of users. Most directly, others have raised concerns about how AR content can obstruct a user's perception of the physical world, e.g., by blocking or otherwise impairing central or peripheral vision.[24]

As a more subtle effect, we find particularly compelling the case of Motion Induced Blindness[25] that can be demoed online at http://www.michaelbach.de/ot/mot-mib/. In this perceptual phenomenon, perfectly visible motion signals disappear from sight for a significant interval -- long enough, for example, to impair the estimate of a trajectory of an object moving at us. In the AR context, the threat may be represented by an introduced motion pattern that causes relevant moving objects in the physical world to disappear. An AR system could use eye tracking to, for example, ensure that the induced pattern was always in a certain region of the person's visual perception. In another case, motion presented in the periphery of the visual field could be misused to draw a user's attention to that region, and away from more important information in the center of the visual field.[26]

---

[23] D. C. Burr. "Motion Perception, Elementary Mechanisms." In *The Handbook of Brain Theory and Neural Networks*, Second Edition. MIT Press, 2004
[24] E. E. Sabelman and R. Lam. "The Real-Life Dangers of Augmented Reality." In *IEEE Spectrum*, 2015.
[25] Y. S. Bonneh, A. Cooperman, and D. Sagi. "Motion-induced blindness in normal observers." In *Nature*, 411(6839), 798-801, 2001.

While AR is new, the threats raised here have deep historical roots. For example, in the physical security space, it is well known that adversaries -- in particular, pickpockets -- can exploit a person's tactile perception system to achieve a malicious goal. Namely, one common technique that pickpockets use is to bump the victim somewhere on their body, drawing the victim's attention to the bump, and then steal from the victim's pocket at the same time.

# 3. A Computer Security Perspective

In this paper, we take a computer security perspective on AR-based sensory and perceptual risks like those exemplified by the above case studies. Before presenting our framework for evaluating such possible threats in the next subsection, we step back to discuss *how* such risks *might* manifest into attacks against real AR users. We stress, however, that such threats have not yet manifested today -- rather, our goal is to look ahead and anticipate such issues before they arise in real AR technologies and applications.

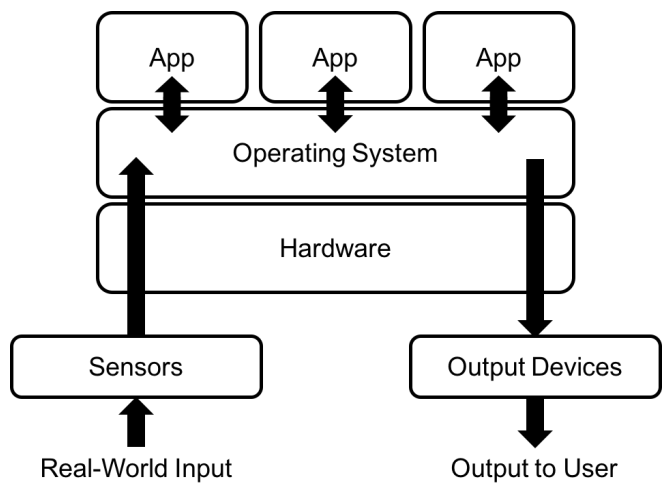

*Figure 2: Conceptual AR Platform.*

We first describe a possible AR platform at a conceptual level, which consists of hardware and software components. The hardware includes both sensors (e.g., RGB and depth cameras, accelerometers, gyroscopes and magnetometers, microphones, eye trackers, biosensors, thermal sensors) and a display, which may take the form of (for example) a head-mounted display, an automotive windshield, or a smartphone screen. The software includes both the platform's operating system (OS) as well as any applications running on top of it. These applications process sensor input and produce output in the form of digital (audio, visual, or haptic) content to be overlaid on the user's perception of the physical world. Applications may be written by developers trusted by the platform designers (e.g., other teams within an automotive company) or by untrusted third-party developers (e.g., who make their applications available in an App Store, similar to today's smartphone ecosystems).

---
[26] A. P. Hillstrom and S. Yantis. "Visual motion and attentional capture." In *Attention, Perception, & Psychophysics*, 55(4), 399-411, 1994.

Our focus in this paper is on the potential sensory and perceptual threats that can arise from the output produced by AR technologies -- for example, output that triggers epileptic seizures or motion-induced blindness. Possible intentional or unintentional "attackers" include the AR platform or OS itself, installed applications, or other users (e.g., when using collaborative applications that allow one user to create content that another user can see). Such "attacks" can occur either by *accident* -- e.g., due to bugs in a trusted application, or because two AR applications interacted in an unexpected way -- or due to *explicit malice* -- e.g., when a user accidentally installs a malicious (or compromised) application from the app store.

We again stress that our exploration in this paper is of *potential* threats. It is unlikely that most of these concerns will manifest in the AR technologies commercially available today; rather, our goal is to encourage current and future AR platform designers to think about these issues in advance, before AR technologies become interesting targets for real adversaries. We also highlight that computer security is fundamentally about risk management. No system can be perfectly secure against all possible adversaries, and system designers must balance security against other goals, including functionality and usability. Just because an attack might be possible does not mean that the risk of this attack is high. For example, the risk of a given attack may be low because it requires a sophisticated, highly motivated adversary with significant resources, or because the potential victim and/or harm is not of sufficient interest to the adversary.

# 4. Framework for Sensory and Perceptual Risks

In this section, we draw on the above case studies, and our computer security perspective, to develop a conceptual framework for classifying potential perceptual and physiological risks and threats in AR. We then use this framework to evaluate a set of concrete examples in the subsequent section (see Table 1).

Our goal in developing this framework is to lay the groundwork for future work that studies and defends against these types of risks, e.g., to help such work anticipate risks and prioritize those that are more likely to arise in practice and/or that will result in greater harm to users.

We define our framework along the following axes:

**Impact.** What aspect of the user's perception and cognition is impacted by the attack? Possible impacts include manipulating a user's perception (visual, acoustic, or haptic), motor control, attention, higher-level decision making, beliefs, emotions, or memory. In this paper, we focus primarily on *sensory and perceptual* risks (as in the case studies above), rather than threats to higher-level brain functions like memory or decision-making. Future work should also consider such higher-level threats in more depth.

**Threat Vector.** How is the attack delivered to the user? Malicious or harmful stimuli may be presented in visual, audio, or haptic form. All of our case studies above involve threats via visual input, but similar risks can arise via other input modalities (e.g., adaptation to auditory stimuli).

**Time to Impact.** How long does it take for the impacts of the attack to take effect? For example, for some type of adaptation to occur one needs an "adapter stimulus" that lasts at least 30 seconds. In some cases, an attacker may deliberately deploy an attack over a longer period of time to reduce the likelihood that the user notices the attack, e.g., by gradually changing the delta between the real world and the augmented world over time.

**Longevity of Impact.** How long do the effects of the attack last? For example, some attacks may have only short-term sensory or perceptual impacts (e.g., temporary motion blindness), while others (e.g., the McCollough Effect) can last up to months; some effects, particularly if applied during critical periods of brain development, may be permanent. For example, in a future when these technologies are pervasive throughout the population, malicious content could deprive young users of certain sensory experiences and perturb their neural development and plasticity (e.g., permanently affecting their ability to perceive certain types of visual information).[27] Any effects that last beyond the application of the attack stimulus are particularly dangerous because they mean that users -- even if they detect an attack -- cannot return themselves to a normal state simply by removing or disabling the AR device.

**Vulnerable Population.** Which users, under which conditions, are vulnerable to the attack? For example, attacks targeting people vulnerable to epileptic seizures are effective only against such users, while a more general population is vulnerable to motion-induced blindness. The population vulnerable to a given attack may vary by age (e.g., children in a critical developmental period[28]), vision, genetic condition, profession (e.g., military or medical applications of AR), current physiological or psychological state (e.g., tired or drunk), etc. And further, will the attacker be able to know whether a potential victim is in the vulnerable population *before* initiating the attack (e.g., by collecting data from the AR device's biosensor or using auxiliary data about the user)?

---

[27] D. Ellemberg, T. L. Lewis and D. Maurer. "Repeated measurements of contrast sensitivity reveal limits to visual plasticity after early binocular deprivation in humans." In *Neuropsychologia*, 44(11): 2104-12, 2006.

**Attack Source.** What is the source of the threat, i.e., what actor is the cause of the threat? Attacks may result from a compromised operating system, from malware or vulnerable/compromised applications, from third-party content (e.g., ads) within an application, from other users (e.g., in a collaborative AR setting), or due to errors, bugs, or accidents on the part of OS or application developers.

**Attack Certainty.** We might have high confidence that an attack will be effective given current knowledge in computer security and neuroscience. Others attacks one might posit as being possible but would require additional (ethical) experimentation in order to fully gauge their feasibility and impact.

**User Awareness.** Can users be aware that the attack is occurring, perhaps early enough to mitigate the attack? Can users be aware that the attack *has* occurred? For example, users may not be aware that they have failed to perceive something due to motion-induced blindness, while they will certainly be aware of an epileptic seizure.

**System Awareness.** Can the ongoing or completed attack's effects on the user be observed externally, e.g., by biometric sensors incorporated into the AR platform? For example, tracking a user's eye movements, heart rate, or other physiological responses may give indication that an attack is occurring. Note that this capability could be used both defensively (i.e., to detect and alert users to an attack) or offensively (to determine attack success and/or optimize an attack).

# 5. Survey of Potential Sensory and Perceptual Threats from AR

Table 1 (see Appendix) surveys the broader space of potential sensory and perceptual threats in a "cheat sheet" format for practical access, and demonstrates how each threat can be evaluated according to the framework developed in the previous section. This table also includes our case studies from above, and provides major references to expand on each topic.

---

[28] J. O. Bailey and J. N. Bailenson. "Immersive virtual reality and the developing child." In P. Brooks and F. Blumberg (Eds.), *Cognitive Development in Digital Contexts* (p181-200). Elsevier, 2017.

# 6. Discussion and Future Directions

Finally, we step back and discuss key directions for future work suggested by our exploration.

**Experimental Evaluation.** In this paper, we have conceptually considered potential sensory and perceptual harms due to AR, based on our understanding of prior literature in perceptual psychology and neuroscience. It will also be critical to experimentally validate, and further evaluate according to our proposed framework, potential AR-based threats to human perception, physiology, and cognition. Experimentally evaluating the differences between AR and other avenues of perceptual attack is also important: for which perceptual attacks does AR provide a unique opportunity for attackers (e.g., due to AR's immersive nature or wide field-of-view), and how effective can these attacks be in practice? Any such experimentation must be done both ethically and safely.

**Additional Neuro-Cognitive Threats**. In our case studies, as well as Table 1, we have focused primarily on perceptual threats, as an initial exploration of this space. Future work must also consider higher-level effects on the human neuro-cognitive system, including effects on cognition, decision-making, emotion, and memory. Our goal in proposing the evaluation framework above is for other AR, neuroscience, and/or computer security researchers to apply and expand this framework as well as Table 1.

**Future AR Technologies.** As AR technologies continue to advance, they may pose additional threats or increase the potential risks or likelihoods of the threats we have identified. For example, future AR platforms may include more and improved biosensors like eye tracking , pupillometry, galvanic skin response (GSR), electroencephalogram (EEG), heart rate; they may also have improved output capabilities (e.g., wider field of view, tactile feedback), as well as a potential for human-like, photo-realistic digital assistants, which project the risk of brain "hacks" to higher level functions like decision-making.

**Defenses.** Our ultimate goal in raising and exploring potential AR-based risks to the human brain is to prevent such risks from manifesting in the next generation of AR technologies. By anticipating and understanding these risks, our hope is that AR technology designers can take them into account and proactively develop defenses against them *before* such risks arise in practice. For example, possible defensive directions include:
- Equipping an AR platform or operating system to monitor for physiological or other signals (e.g., heart rate) that can detect a user's abnormal reactions and flag potential attacks when they begin taking effect.
- Designing the AR platform or operating system to mediate and validate any output that applications attempt to generate, to detect signatures for potential (accidental or intentional) attacks. Prior work in the computer security community has already considered the role of the AR's operating system in managing output from applications,[29] but it has not studied in depth how to do so to prevent attacks of the sensory and perceptual nature we discuss in this paper.
- Vetting AR applications submitted to an App Store via static and/or dynamic analysis, to detect potential sensory or perceptual risks and/or violation of app development guidelines, and proactively removing from the App Store applications that are found to be problematic once already deployed.
- Rules about the nature of content that is safe for the perception of *other* users when AR is used collaboratively among multiple users. (For example, as when a group of people share a meal and must know others' dietary restrictions, one user's AR device may take into account the fact that another user is predisposed to photosensitive epilepsy before sharing virtual content.)

# 7. Conclusion

Rapidly advancing AR technologies are in a unique position to directly mediate between the human brain and the physical world. Though this tight coupling presents tremendous opportunities for human augmentation, it also presents new risks due to potential adversaries, including AR applications or devices themselves, as well as bugs or accidents. In this paper, we have begun exploring potential risks to the human brain from augmented reality. Our initial focus has been on sensory and perceptual risks (e.g., accidentally or maliciously induced visual adaptations, motion-induced blindness, and photosensitive epilepsy), but similar risks may span both lower- and higher-level human brain functions, including cognition, memory, and decision-making. Though they have not yet manifested in practice in early-generation AR technologies, we believe that such risks are uniquely dangerous in AR due to the richness and depth with which it interacts with a user's experience of the physical world. We propose a framework, based in computer security threat modeling, to conceptually and experimentally evaluate such risks. The ultimate goal of our work is to aid AR technology developers, researchers, and neuroscientists to consider these issues *before* AR technologies are widely deployed and become targets for real adversaries. By considering and addressing these issues now, we can help ensure that future AR technologies can meet their full, positive potential.

# Acknowledgements

This work was supported in part by the National Science Foundation under Award CNS-1651230.

---

[29] K. Lebeck, K. Ruth, T. Kohno, and F. Roesner. "Securing Augmented Reality Output." In *Proceedings of the 38th IEEE Symposium on Security and Privacy*, 2017.

# Appendix: Survey of Potential Sensory and Perceptual Threats from AR

*Table 1: Taxonomy of Potential Sensory and Perceptual Threats from Augmented Reality*

| Threat Vector | Description of Impact | Attack Vector | Time for Delivery | Time to Impact | Longevity of Impact | Vulnerable Population | Attack Certainty | User Awareness |
|---|---|---|---|---|---|---|---|---|
| *Temporal modulations*[1] | Deliver high contrast stimuli at a high temporal frequency to generate photosensitive epilepsy and discomfort. | Visual | Delivered over <1 minute. | Seconds, Minutes | Seconds, Days | Predisposed people, more frequent in young | Low (conditional on pre-disposition) | Yes |
| *Visual fatigue*[2] | Repetitive or prolonged presentation of certain visual stimuli can generate visual distress and fatigue. | Visual | Delivered over prolonged intervals and repetitive stimulation. | Minutes, Hours | Minutes, Hours | General. High risk for machine operators | Medium | No |
| *Adaptation and illusions*[3] | Present stimuli in any modality to adapt those senses so they are more sensitive to the opposite of the stimuli. See "Case Study 2". | Visual, acoustic, haptic | Delivered over various durations, from single flashes to minutes or hours. | Seconds, Minutes, Hours | Seconds to Days: Effect of variable duration sometimes correlated to the adapting phase duration. | General. High risk for machine operators | High | No |
| *Motion*[4,5] | Deliver motion signal unexpectedly in target areas of the field of vision to create attentional capture and distraction, disorientation, vection, and stimulus masking. | Visual, acoustic | Delivered over a few seconds. | Seconds | Seconds to Days: Effect lasting over tens of seconds or less | General. High risk for machine operators | High | Conditional |
| *Attentional system*[6] | Capture attention by popping up high salient stimulus as distraction. | Visual, acoustic | Delivered over short or very short intervals. | Seconds | Seconds: Effect is generally instantaneous or short term (but secondary effects may be long lasting or permanent). | General. High risk for machine operators | High | Conditional |
| *Binocular disconnections*[7] | Present incoherent stimuli in each eye to cause double image (diplopia), distance mis-estimations. | Visual | Delivered in short or long duration. | Seconds | Seconds, Minutes | General. High risk for machine operators | Low | Yes |
| *Binocular rivalry*[8] | Presenting different stimuli in two eyes can cause suppression of the information in one eye, thus cause disruption of stereo vision, disorientation, discomfort, priming, nausea. | Visual | Delivered over short or very short intervals. | Seconds, Minutes | Seconds, Minutes | General. High risk for machine operators | High (conditional on intact stereo vision) | Conditional |
| *Sound / acoustic attacks*[9] | Deliver sound stimuli that may contain certain spatial cue or a certain sound frequency for a long period of time to cause temporal biases, spatial biases, tynnitus, hyperacusis. | Acoustic | Delivered over short or very short intervals. | Seconds | Seconds, Months | General. High risk for machine operators | Medium | Conditional |
| *Multisensory disconnections*[10] | Present inconsistent stimuli from different sensory channels and cause confusion that includes spatial and temporal mislocalizations, disorientation, and Simulator Sickness (nausea). | Visual, acoustic, haptic | Delivered over short or very short intervals. | Seconds, Minutes | Seconds, Minutes | General. High risk for machine operators | High | No |

| | | | | | | | | | |
|---|---|---|---|---|---|---|---|---|---|
| *Sensory motor conflicts*[11] | Introduce conflicting visual and motor cues to cause loss of accuracy of targeted movements such as reaching/ grasping. | Motor | Delivered in short or long duration. | Minutes | Minutes, Hours | General. High risk for machine operators | High | | |
| *Delusions / Hallucinations*[12] | Overlay photorealistic images on top of real world scenes which distort the perception of reality. | Visual, acoustic | Delivered over prolonged intervals and repetitive stimulation. | Seconds, Minutes | Minutes, Lifetime: | General | Low (conditional on pre-disposition) | No | |

**Table References:**